\documentstyle[aps,preprint,psfig,prb]{revtex}
\begin{document}

\draft

\title{Fano effect in the $a-b$ plane of Nd$_{1.96}$Ce$_{0.04}$CuO$_{4+y}$:
evidence of phonon interaction with a polaronic background} 
\author{S. Lupi, M. Capizzi, P. Calvani, B. Ruzicka,\cite{Ruzicka} P. Maselli, 
P. Dore, and A. Paolone\cite{Paolone}} 
\address{Istituto Nazionale di Fisica della Materia - Dipartimento di Fisica, 
Universit\`a di Roma ``La Sapienza'',}
\address{Piazzale A. Moro 2, I-00185 Roma, Italy}
\date{\today}
\maketitle

\begin{abstract}
Reflectance measurements in properly selected samples of 
Nd$_{1.96}$Ce$_{0.04}$CuO$_{4+y}$ with different carrier concentrations provide
a firm evidence of Fano antiresonances for the four transverse optical $E_u$ 
phonons in the $a-b$ plane.
A fit of the whole optical conductivity has been performed, without any
previous subtraction of the electronic background. Those fits, as well as the
dependence on temperature of the renormalized phonon frequencies, show that the 
electronic continuum is provided by the same polaron band which characterizes 
the optical conductivity of all the insulating parent compounds of high T$_c$ 
superconductors. The relevance of these results for the metallic and
superconducting phase of cuprates is discussed.
\end{abstract}


In 1961, U. Fano quantitatively accounted for the so called Fano antiresonance,
i.e., the asymmetric line shape of the $2s2p^1P$ resonance of He observed in 
electron inelastic scattering experiments.\cite{Fano61} In that work, Fano 
revised an earlier qualitative interpretation\cite{Fano35} of the spectra of 
He and other rare-gases. Fano pointed out that such effects are
expected whenever a set of discrete states is mixed with a continuous 
spectrum. Thereafter, ``Fano profiles'' have been 
observed in a number of spectra, including those where phonon discrete
states interact with a continuum background due to itinerant charges.
In high critical temperature superconductors (HCTS), phonon Fano profiles have 
been detected with the electric field of the radiation polarized along the 
$c$-axis, either in\cite{Macfarlane87,Cooper88,Thomsen88,Feile88} 
YBa$_2$Cu$_3$O$_{7-\delta}$, and in\cite{Zibold92} 
Bi$_2$Sr$_2$CaCu$_2$O$_8$. Moreover, in the 
400-600 cm$^{-1}$ region of the reflectivity spectra of 
Pb$_2$Sr$_2$LCu$_3$O$_8$ (L = Y, Dy, Eu, 
Nd, and Pr), phonon lineshapes have been reported to become 
asymmetric for increasing doping.\cite{Reedyk92} These results have been 
explained in terms of an interaction of Raman-active modes along the $c-$axis, 
made infrared-active by some symmetry-breaking 
potential, and an electronic continuum which develops with doping in the 
mid infrared. As far as we know, no evidence of Fano lineshapes has been 
reported for the four infrared active $E_u$ phonons of the $a-b$ plane. Here
indeed, the optical phonons are shielded by the carriers which form below 
$T_c$ a fluid of superconducting pairs.

In order to look for Fano resonances in the $a-b$ plane of a
cuprate, we have selected a strongly doped, non-metallic system, where 
the existence of a considerable electron-phonon coupling is ensured by 
previous observations of
polaronic effects in the $a-b$ plane reflectivity. It is the case of 
Nd$_{2-x}$Ce$_x$CuO$_{4-y}$ (NCCO), where
an electronic continuum detected at $\sim$ 1000 cm$^{-1}$ has been explained 
in terms of a polaron-band $d$, whose strength increases with 
either $x$, see Ref. \onlinecite{Lupi92}, or $y$, see 
Ref. \onlinecite{Calvani96}. 
In order to discriminate the effect of the charges added to the $a-b$ plane 
from possible effects due to the chemical doping, two single crystals with 
the same Ce doping but different carrier concentration have been 
measured, MN10 and MN29. The former is as-grown, the latter is enriched in 
oxygen. From the present spectra one checks that the effective 
number of carriers available in the charge-transfer gap for MN10 is $\sim$ 
1.5 times that for MN29.\cite{phys-bb} 
In the following it will be shown that asymmetric lineshapes are found in 
both samples for the four $E_u$ phonons of the $a-b$ plane, which can be 
explained in detail by a Fano interaction between the phonons and an 
electronic background.\cite{Davis77} This latter is not the weak Drude term
eventually present in these semiconducting compounds, but {\it the polaron 
$d$ band peaked at $\approx$ 1000 cm$^{-1}$}. This conclusion is based on the 
position of the dips in the Fano lineshapes and on the displacement at low 
temperature of the Cu-O stretching frequency. This phonon shift is 
negligible in MN29, where the $d$ 
band changes weakly with temperature, increases to 7 cm$^{-1}$ in MN10 where 
at low $T$ the polaron band undergoes a dramatic transfer of spectral weight 
towards lower energies. 

The experimental apparatus has been described in detail
elsewhere.\cite{Calvani96} Data have been collected from 20 (MN29) or
70 (MN10) through 18,000 cm$^{-1}$ by use of a Michelson interferometer. The 
real part of the optical conductivity $\sigma(\omega)$ has been obtained from 
Kramers-Kronig transformations. The reflectivity data have been extrapolated 
to zero frequency either by using a Hagens-Rubens formula or by a 
Drude term obtained by fitting a Drude-Lorentz model for the complex dielectric
function $\widetilde\epsilon(\omega)$ to the $R(\omega)$ data.\cite{Calvani96}
The transformed $\sigma(\omega)$, for $\omega\ge 70 (20)$ cm$^{-1}$, has been 
found to be independent of the approach used. For $\omega\ge 18,000$ 
cm$^{-1}$, $R(\omega)$ has been extrapolated up to $\omega$ = 320,000 cm$^{-1}$
by using the values at 300 K reported by Tajima {\it et al}\cite{Tajima} and, 
at higher frequencies, by extrapolating $R(\omega)$ with a $\omega^{-4}$ law. 

The low-energy reflectance $R(\omega)$ of 
samples MN10 and MN29 is reported in Fig. 1 for $T$=20 K and $T$=300 K.
$R(\omega)$ is strongly $T$-dependent in both samples and lower in 
MN29 than in MN10, consistently with the lower number of extra charges in the 
former crystal. In both samples, the room temperature
reflectivity is dominated by the four $E_u$ phonons and by a broad but clear
peak at $\alt$ 1000 cm$^{-1}$, the $d$ band. In the low-$T$ spectra, several
additional modes add to the extended $E_u$ phonons. They are the IRAV observed 
in the same compound when doped by oxygen vacancies.\cite{Calvani96}
     
The low-energy part of the optical conductivity, obtained as described above, 
is reported in Fig. 2 for the same samples and temperatures of Fig. 1. 
The room temperature phonon lines of both samples (in particular the 
bending and stretching Cu-O modes at $\sim$ 300 and $\sim$ 515 cm$^{-1}$, 
respectively) deviate appreciably from the ordinary Lorentzian shapes. They 
display a strong 
asymmetry, with a dip on the low energy side which indicates an interaction 
with a continuum at higher energy.\cite{Fano61} The closest band with such
features is the $d$ band. In order to verify if really {\it 
the $E_u$ phonons interact with the polaronic background}, 
the optical conductivity from 70 (20) to 18,000 cm$^{-1}$ has been fitted 
in terms of a complex dielectric function given by

\begin{equation}
\tilde\epsilon(\omega) =  
\tilde\epsilon_D +
\tilde\epsilon_F +
\tilde\epsilon_{lm}+
\tilde\epsilon_d +
\tilde\epsilon_{MIR} +
\tilde\epsilon_{CT} +
\epsilon_\infty,
\end{equation}

\noindent
i.e., by a (weak) Drude-like term, Fano resonances, local modes, the polaron 
$d$ band, the mid-infrared band MIR, the charge-transfer band, and finally 
by all higher-energy oscillators which are represented by 
$\epsilon_\infty$ (see Ref. \onlinecite{Lupi92} for details on most of 
these terms). These contributions are all given by $j$ Lorentzians, but for 
$\tilde\epsilon_F$, with strengths, peak frequencies, and damping factors
given by $S_j$, $\omega_j$, and $\gamma_j$, in the order. The overall 
contribution to the dielectric function of an oscillator interacting 
with an electronic continuum has been first estimated
in Ref. \onlinecite{Davis77}. This estimate has been never used before
to fit the far-infrared optical conductivity of HCTS, or of other ionic 
compounds, at least to our knowledge. Fano lineshapes are usually fitted,
instead, after an arbitrary subtraction of the electronic background.
Following Ref. \onlinecite{Davis77}, one has 
$\tilde\epsilon_F=\sum_k\tilde\epsilon_{kF}$, with

\begin{equation}
\epsilon_{1,kF}(\omega) = 
                          R_k\gamma_k\left(
                          {{(q_k^2-1)(\widetilde\omega_k-\omega)+2q_k\gamma_k}
                        \over{\gamma_k^2+(\omega-\widetilde\omega_k)^2}}\right),
\end{equation}

\noindent
and 

\begin{equation}
\epsilon_{2,kF}(\omega) = 
                   R_k\left({{(q_k\gamma_k+(\omega-\widetilde\omega_k))^2}
                   \over{\gamma_k^2+(\omega-\widetilde\omega_k)^2}}-1\right), 
\end{equation}

\noindent
where $R_k$ are scale parameters which include the transition rates to 
the excited states and $q_k$ are the Fano-Breit-Wigner parameters, an inverse 
measure of the electron-phonon interaction for the $k$-th phonon.\cite{Fano61} 
The dependence of $R_k$, $q_k$ and $\gamma_k$ on
energy is assumed to be smooth and is therefore neglected in the model.
$\widetilde\omega_k$ is the renormalized frequency of the $k$-th phonon.

The fits obtained in this way are excellent in both samples at all 
temperatures, as shown by the close match in Fig. 2 between the theory (full 
lines) and the data (dots) in the case of 300 K and 20 K. The fit parameters 
are given in Table I for samples MN10 and MN29, together with those of the 
polaron 
band. For all $E_u$ modes of sample MN10, $q$ is roughly constant for $T \leq 
200$ K, while it decreases by a factor of 3-4 at room temperature. These values 
are similar, but for the sign, to those reported in 
YBa$_2$Cu$_3$O$_{7-\delta}$ for the Raman mode at 312 cm$^{-1}$ interacting 
with an electronic background. Therein, $q$ is equal to -4.3 at room 
temperature\cite{Thomsen89} and ranges from -6 to -2 for temperatures varying 
between 4 K and 300 K.\cite{Feile88} In that case,
however, Raman-active $c$-axis phonons were claimed to interact with a Drude
continuum at zero frequency, hence the negative sign for $q$. 

The general optical properties of NCCO in the midinfrared are described 
in detail elsewhere.\cite{phys-bb} However, the dependence of the $d$ band 
on doping and temperature (Fig. 2) will be examined
here in relation to the present discussion of the phonon Fano lineshapes. 
As a whole, the behavior of the $d$ band for a 0.04 Ce doping follows that 
already observed in ligthly oxygen-deficient  NCO.\cite{Calvani96} 
In the compensated sample MN29 (see Fig. 2 and Table I), the peak energy 
of the $d$ band is 
$E_{max}$=1100 cm$^{-1}$ at 300 K and slightly decreases to 850 cm$^{-1}$ at 
20 K. In the as-grown MN10, $E_{max}$=900 cm$^{-1}$ at 300 K, and drops to 
480 cm$^{-1}$ at 
20 K. In Table I one can also notice that in sample MN10 the highest frequency 
(stretching) mode hardens for decreasing temperature, by an amount much 
larger than the experimental uncertainty of $\sim \pm$ 0.5 cm$^{-1}$, while 
the energy of the other phonons remains constant. On the contrary, in sample 
MN29 no appreciable dependence on $T$ 
has been detected either for the energies of all the four $E_u$ phonons, 
{\it including the stretching mode}, or for their Fano parameters. For all
MN29 phonons $q \approx 10$ but for the bending mode, where $q \sim 3$.  
The dependence on $T$ reported above for the two 
samples confirms the conclusion based on the position of the dip in the
asymmetrical phonon lineshapes, namely that the {\it the phonons 
interact via Fano with the polaron $d$-band}. In fact, for 
decreasing $T$ the $d$ band shifts 
to low energy by a large amount in sample MN10, only slightly in sample MN29.
Moreover, in the former sample the difference between $E_{max}$ and 
the peak energy of the phonon changes drastically for the highest energy 
phonon, which hardens as predicted by the theoretical 
model.\cite{Fano61,Zibold92} 
Also the dependence on $T$ of the Fano parameter $q$, inversely related to the 
strength of the electron-phonon interaction, is consistent with the behavior 
with $T$ of the polaron band in the two samples. In fact, 
from $E_p$ = $g^2/\omega^*$, where $E_p$ is the binding polaron energy, $g$ is 
an average electron-phonon-interaction parameter, and $\omega^*$ is an average
interacting-phonon frequency. The latter is $\sim$ 210 cm$^{-1}$ in NCO from 
Ref. \onlinecite{Calvani96}, in a Lorentzian band model, and from 
$E_{max}$=2$E_p$ ($\approx$4$E_p$) for a small (large) polaron,\cite{Emin93} 
one obtains 
$g$=$(A\omega^* E_{max})^{1/2}$, with A=0.5 ($\approx$0.25). In a small polaron 
approach, therefore, the softening of $E_{max}$ in sample MN10 is related to 
a parallel decrease of $g$ (from 300 cm$^{-1}$ to 220 cm$^{-1}$ for $T$ going 
from 300 K to 20 K) and an ensueing increase of $q$ in the same temperature
range (in a large polaron approach, $g$ changes from 220 cm$^{-1}$ 
to 160 cm$^{-1}$). In sample MN29, instead, the energy of the polaron band has 
a smaller relative change with $T$ while $q$ is roughly constant.
If one extrapolates these results to even higher doping (where direct 
observations of phonon and $d$ bands are prevented by shielding effects) 
one may infer that the insulating-to-metal transition
observed in NCCO for $x$=0.13 could be due to a collapse of the binding
energy of the polaron, which could turn from a partially localized
quasiparticle to a completely delocalized carrier.

The strength of the electron-phonon interaction can be quantitatively estimated
also from the $T$ dependence of the stretching mode energy in sample MN10, 
{\it the only mode which shows a sizable $T$ dependence}. It has to be noticed 
that this $T$ dependence cannot be ascribed to changes in the lattice 
parameters since the same stretching mode is roughly $T$ independent 
in sample MN29. 
By using the Green's function formalism,\cite{Zibold92} one obtains for the 
renormalized phonon frequencies

\begin{equation}
\widetilde \omega_k = \omega_k 
                      \left(1 - {{g^2_k(\omega^2_e-\omega^2_k)/2)}
                      \over{(\omega^2_e - \omega^2_k)^2 + (\gamma_e\omega_k)^2}}
                      \right),
\end{equation}

\noindent
where g$_k$ are the coupling constants for the interaction of the $k$-th 
phonon of energy $\omega_k$ with an electronic oscillator with resonance 
frequency and damping constant given by $\omega_e$ and $\gamma_e$, 
respectively. It is worth noticing here that the coupling constants $g_k$ do 
not necessarily coincide with the
electron-phonon interaction parameter $g$, for at least two reasons. First, $g$
is an effective parameter averaged over the different local
modes\cite{Calvani96} or phonons which build up the polaron band. Second, the
value of the interaction
which leads to the formation of a polaron band may differ from that between 
the same band and extended phonons which results into the Fano resonances.
Therefore, in evaluating the renormalized phonon frequencies 
$\widetilde \omega_4$ from Eq. 4 we take $g_4$ as a fitting parameter which, 
for sake of simplicity, is assumed to be temperature independent.
The bare phonon frequency $\omega_4$ is the second fitting parameter, while the
values of $\omega_e$ and $\gamma_e$ entering Eq. 4 are those obtained 
at each temperature by the Lorentz-Fano fit of the optical conductivity and 
reported in Table I. The dependence on temperature measured for the 
stretching-mode energy $\widetilde \omega_4$ is reported by dots in Fig. 3,
together with the theoretical estimate of 
$\widetilde \omega_4$ as obtained for $g_4 \simeq$ 150 cm$^{-1}$ and 
$\omega_4 = 514 \pm 1$ cm$^{-1}$. 
The agreement between the model and the experiment is excellent. The value
for $g_4$ is on the same order of that estimated for $g$, thus strengthening
the whole model of interaction between carriers and the polaron band.
The same model, when applied to the other three modes in sample MN10 or to all
the four modes in sample MN29, gives values of the phonon energies which are 
constant within the experimental uncertainty, in agreement with the 
experimental results, provided $g_k \alt 80$ cm$^{-1}$. 

In summary, a clear signature of a Fano interaction between all the transverse 
optical  
$E_u$ phonons and an electronic continuum has been obtained by measuring the 
reflectance of two Nd$_{1.96}$Ce$_{0.04}$CuO$_{4+y}$ samples with
different carrier concentrations. A fit has been made of the optical
conductivity by a formula which accounts for the whole effects of the electron
phonon interaction, both on the discrete and on the continuum states. 
In this way it has been shown that those phonons interact with the same polaron 
band which, as recently reported, characterizes the optical properties of 
the insulating parent compounds of high T$_c$ superconductors. The dependence
on temperature of the renormalized phonon frequencies has been analyzed
and an estimate of the strength of the electron-phonon interaction has been
provided, further supporting the above interaction model. These results
point out the role of polarons in the optical properties of the
insulating parents compounds of HCTS. The increase of the Fano parameter 
for decreasing $T$ observed in the more doped sample and the corresponding 
large softening of the polaron band indicate that the insulator-to-metal 
transition in cuprates could be triggered by a collapse to 
zero frequency of the polaron band. Measurements in more doped samples, as well
as in other cuprates, are needed to confirm this intriguing scenario.

We are indebted to W. Sadowski for providing the samples here investigated, 
to G. Strinati and M. Grilli for helpful discussions and suggestions.


\narrowtext
\begin{table} 
\caption
{Parameters of the Fano fit for the $E_u$ phonons and the electronic continuum 
of crystal MN10 (MN29) at different $T$. Units are cm$^{-1}$, except 
for adimensional $q$ and $R$.}   
\label{Table I}

\begin{tabular}{ccccc}
 
                      &    20 K   &   100 K   &   200 K    &     300 K   \\
\tableline
$\widetilde\omega_1$  & 130 (133) &   130     &   130      & 128 (132)   \\
R$_1$                 &   1 (1.2) &     2     &     2      & 4.5 (0.6)   \\
$\gamma_1$            &   3 (8)   &     4     &     4      &   4 (20)    \\
q$_1$                 &   8 (10)  &     8     &     8      & 1.5  (10)   \\
\tableline                                                           
$\widetilde\omega_2$  & 303 (303) &   303.5   &   303      & 304.5 (304) \\
R$_2$                 & 1.9 (3.9) &     1.7   &     3.7    &   9   (2.7) \\
$\gamma_2$            &  6  (5)   &     4     &     4.5    &   5   (6.5) \\
q$_2$                 &  5  (3.1) &     7     &     3.8    &  1.8  (3.1) \\
\tableline
$\widetilde\omega_3$  & 351 (352) &   350     &   349.5    & 350   (351) \\
R$_3$                 & 2.3 (0.68)&     2.7   &     2.3    & 2.8   (0.48)\\
$\gamma_3$            &  5  (10)  &     5     &     5      &   7   (18)  \\
q$_3$                 &  3  (10)  &     3     &     3      & 2.4   (10)  \\
\tableline
$\widetilde\omega_4$  & 517 (515) &   514     &   512      & 511   (514) \\
R$_4$                 & 2.9 (0.83)&     1.7   &     2      &   6   (0.55)\\
$\gamma_4$            &  8  (7)   &     7     &     8      &  11   (12)  \\
q$_4$                 & 4.5 (7)   &     6.5   &     6      &   2    (7)  \\
\tableline
\tableline
$\omega_e$            & 480 (850) &   500     &   570      &  900 (1100) \\
S$_e$                 &5400 (4800)&  5800     &  6900      & 8000 (5400) \\
$\gamma_e$            & 500 (1100)&   650     &   800      & 1100 (1900) \\
\end{tabular}
\end{table}

\newpage

\begin{figure}
{\hbox{\psfig{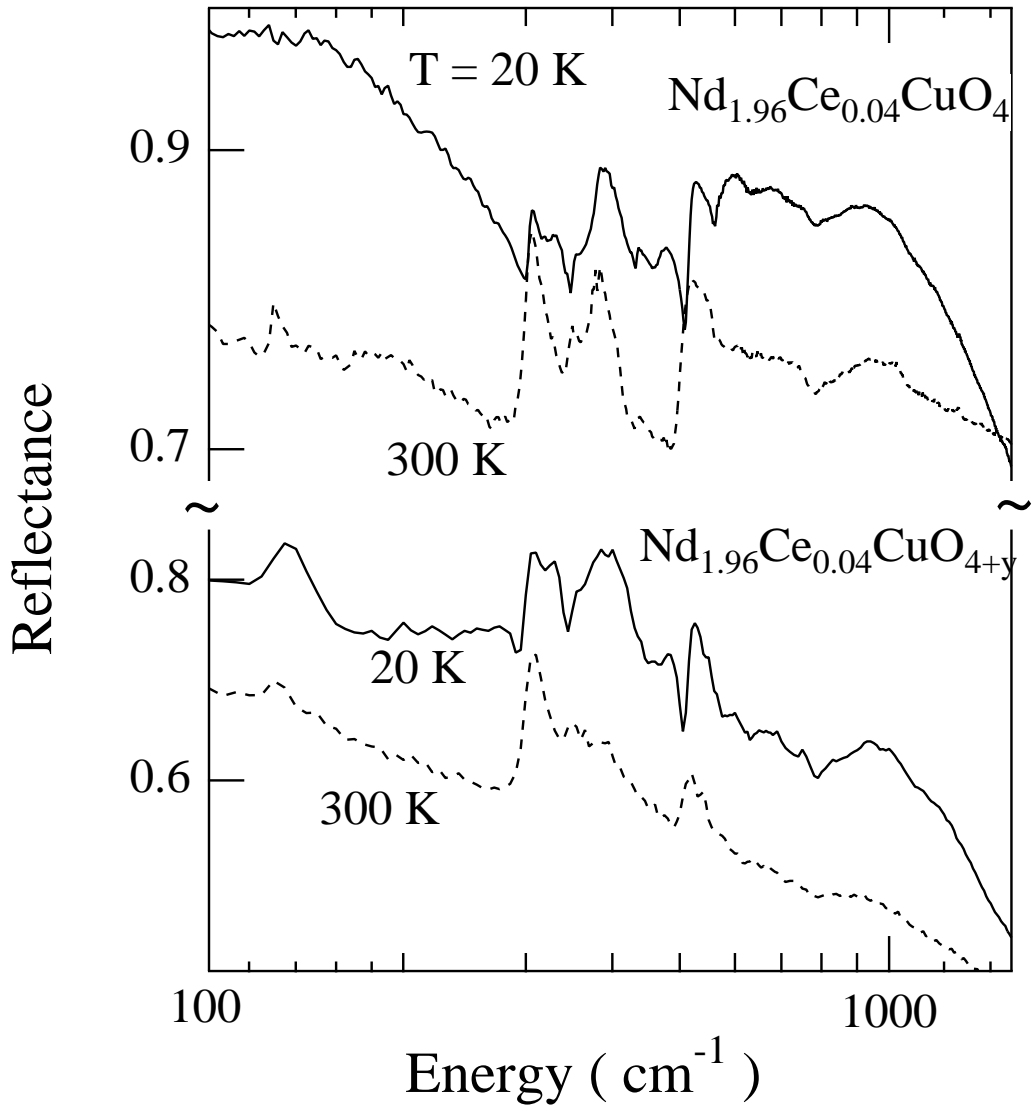}}}
\caption {Far- and mid-infrared reflectance of samples MN10 and MN29, 
from top to bottom, at 20 and 300 K}
\label{fig1}
\end{figure} 

\newpage

\begin{figure}
{\hbox{\psfig{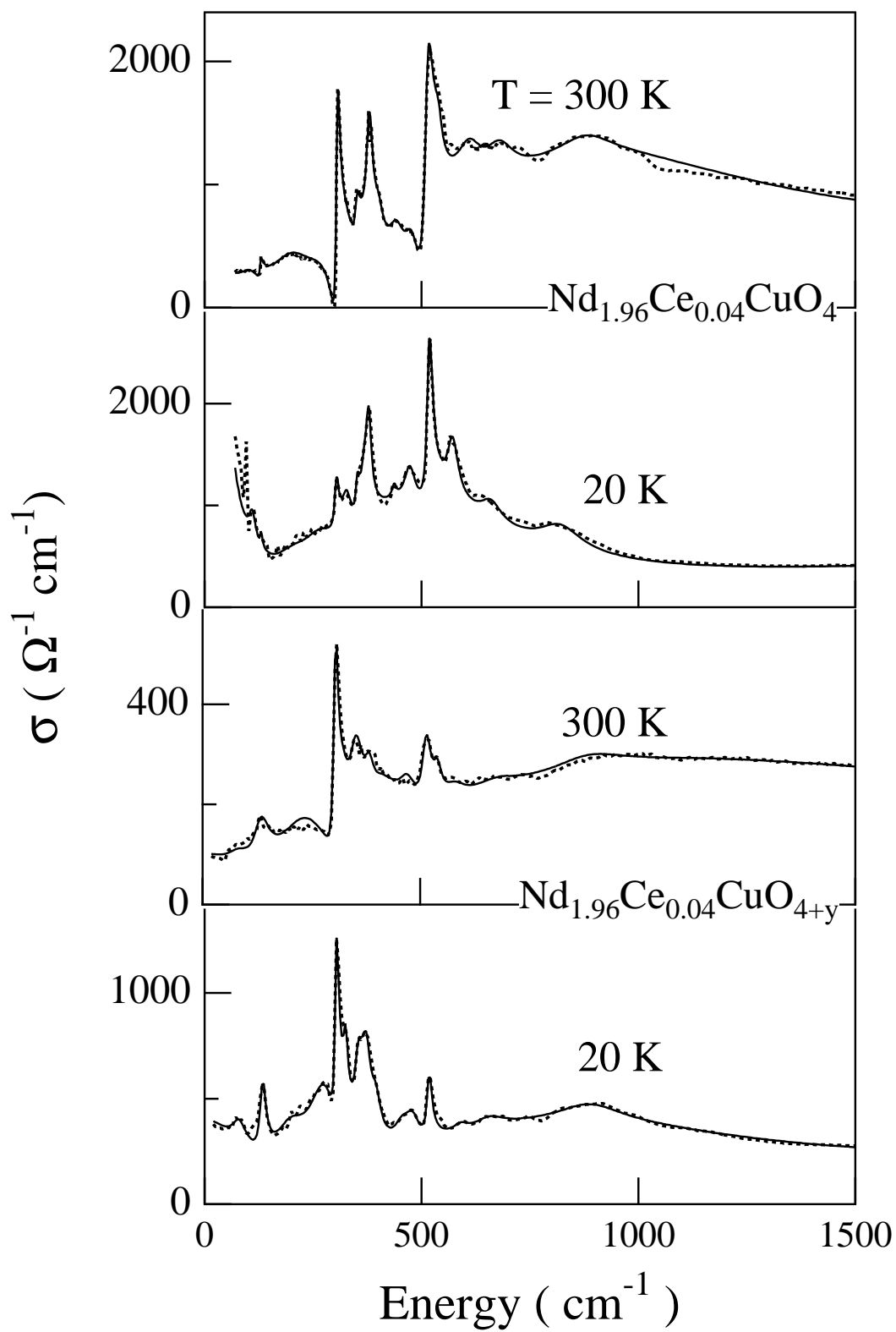}}}
\caption {Far- and mid-infrared optical conductivity of samples MN10 
and MN29, from top to bottom, at 20 and 300 K}
\label{fig2}
\end{figure} 

\newpage

\begin{figure}
{\hbox{\psfig{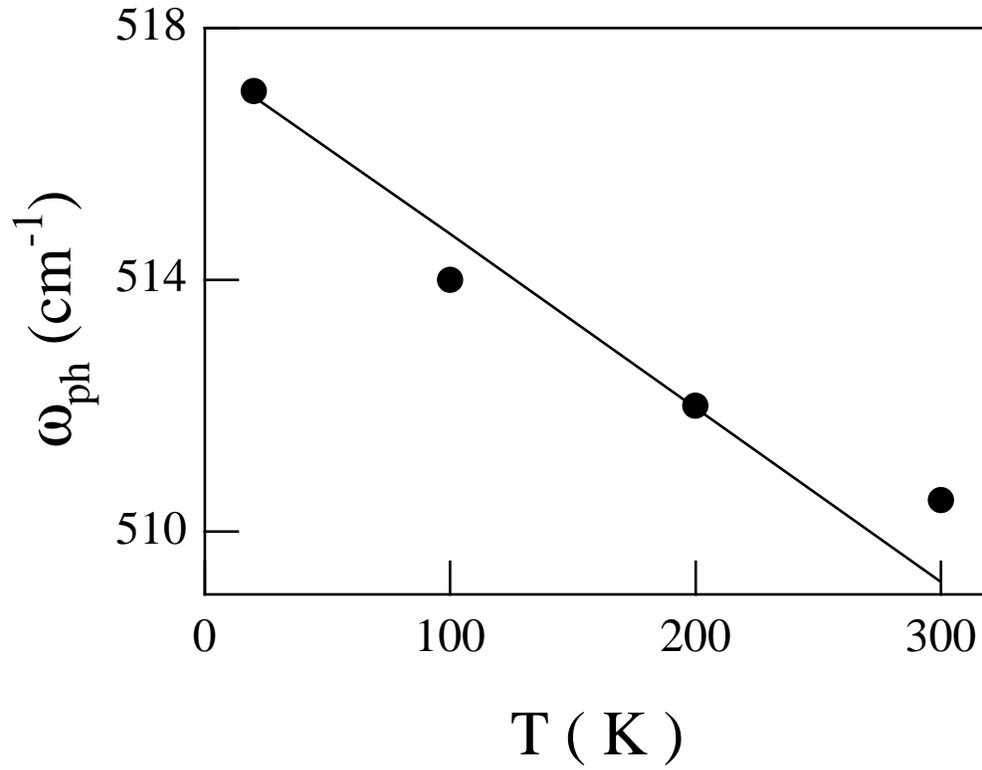}}}
\caption {Frequency of the stretching mode in the most doped 
sample MN10 vs. temperature (dots) and its behavior calculated from Eq. 4 
(solid line).}
\label{fig3}
\end{figure}

\end{document}